\documentclass[twocolumn,showpacs,preprintnumbers,amsmath,amssymb,prc]{revtex4}
\usepackage{epsfig}
\newcommand{\bm}[1]{ \mbox{\boldmath $#1$}  }

\begin{document}

\title{Nuclear $\alpha$-particle condensates: Definitions, occurrence
conditions, and consequences}

\author{N.T. Zinner and A.S. Jensen}
\affiliation{ Department of Physics and Astronomy, University of Aarhus, 
DK-8000 Aarhus C, Denmark}

\date{Received 23 November 2007; revised manuscript received 8 January 2008; published 29 October 2008}

\begin{abstract}
There has been a recent flurry of interest in the possibility of
condensates of $\alpha$-particles in nuclei. In this letter we discuss
occurrence conditions for such states. Using the quantality condition
of Mottelson we show that condensates are only marginally expected in
$\alpha$-particle states. We proceed to demonstrate that few-body
nuclear condensates are ill-defined, and emphasize the conflict
between $\alpha$-localization and $\alpha$-condensate formation. We
also explore the connection between Ikeda diagrams, linear chains, and
Tonks-Girardeau gases. Our findings show that no new information is
contained in the approximations of nuclear states as $\alpha$-cluster
condensates. Furthermore, condensates of more than three
$\alpha$-particles are very unlikely to exist due to couplings to
other degrees of freedom.
\end{abstract}


\maketitle

\paragraph*{Introduction.}
 
The idea of $\alpha$-particles as essential constituents in the
structure of nuclei arises from the small radius, the relatively large
binding and the spin saturation of both neutrons and protons. Attempts
were made in the early days of nuclear physics to construct nuclear
structure from $\alpha$-particles and valence nucleons
\cite{wig37,whe37,wef37}.  In general these attempts were largely
unsuccessful because the nucleon distances within and between
different $\alpha$-particles are comparable.  Thus there are no
compelling reasons for clusterization of nucleons into
$\alpha$-particles.

On the other hand, $\alpha$-cluster models were able to explain many
properties of specific light nuclei \cite{bri66}.  This was
high-lighted by the prediction of a $3\alpha$-structure near threshold
\cite{hoy58} which is crucial for the nuclear synthesis of the heavy 
elements in stars.  This Hoyle state was soon found experimentally
\cite{cook57} and its properties established in a microscopic cluster
model \cite{ueg77} where it was first characterized as a ``gas-like''
structure.  This structure is confirmed in details in numerous
theoretical works \cite{kan98,des02,nef04}.  A radius about 30\%
larger than that of the ground state of $^{12}$C is also reproduced by
use of an approximate wave function consisting of an antisymmetrized
product of three gaussians each containing an $\alpha$-particle
\cite{toh01}.  Also the ground state of $^8$Be was in the same 
approximation described as a gas-like structure of two
$\alpha$-particles \cite{fun02}, and the established structure
\cite{hor70,har72} was again essentially recovered.  The fundamental 
continuum resonance properties were suppressed by use of box-like
boundary conditions effectively supplying a confining external field.

These approximations were presented as novel discoveries of
condensates consisting of two and three $\alpha$-particles.
The last five years have witnessed surprisingly large efforts invested
in  both investigations of the accuracy of the approximation in
\cite{toh01} and extensions to similar simple models for other nuclei
\cite{fun03,yam04}.  The aim seems to be a search for $\alpha$-cluster 
condensates in nuclei. The inspiration is from atomic physics where
Bose-Einstein condensates (BEC) of cold atoms are routinely made and
manipulated by external fields \cite{ing99}.  Related theoretical
investigations are also abundant, see e.g. the review \cite{dal99}.

The concept of BEC is well-defined for macroscopic systems of cold
atoms and molecules. Extensions to self-bound quantum systems with a
small number of particles are not straightforward.  The purpose of
this letter is to discuss the concept of few-body nuclear condensates,
give definitions, compare to cold atomic gases, show the conflict
between localized $\alpha$-cluster models and $\alpha$-condensation,
derive occurrence conditions and investigate consequences.

\paragraph*{Concepts.}

To find $\alpha$-cluster condensates in nuclei two conditions must be
met, i.e. (i) the nucleons must be confined in $\alpha$-clusters, and
(ii) these $\alpha$-particles must form a condensate. Although
$\alpha$-cluster models in general are unsuccessful we shall assume
that (i) holds.  Then to ensure nucleon antisymmetry the $\alpha -
\alpha$ distance must on average be larger than the diameter of the
$\alpha$-particle.

The classical definition of an ideal Bose-Einstein condensate is a
collection of identical particles in the same quantum state.  This
implies that the independent particle model gives an accurate
description which obviously is true for non-interacting particles in
an external field.  For interacting particles the mean-field
description is valid when Mottelsons quantality condition is met
\cite{mot99}, i.e.
\begin{equation} \label{e20}
 \Lambda_{Mot}  =  \frac{\hbar^2}{m c^2_{min}|V_{min}|} > 0.1 -0.2\; ,
\end{equation}
where $m$ is the mass of the particles and $c_{min}$ is the distance
between two particles when the total two-body potential has its
minimum value $V_{min}$.  When $\Lambda_{Mot}$ is small the attractive
potential dominates over the kinetic energy and the particles are
confined to the attractive pockets, i.e. localization or solid
structure. When $\Lambda_{Mot}$ is large the particles can not be
confined by the attraction and the mean-field model is appropriate.

The condition for Bose-Einstein condensates is that the deBroglie
wavelength $\lambda_{dB}$ of the motion must be larger
than the distance $c_{min}$ to the nearest neighbor, i.e.
\begin{equation} \label{e30}
 1< \Lambda_{bec} \equiv \frac{\lambda_{dB}}{c_{min}} = 
 \frac{2 \pi \hbar}{c_{min} \sqrt{2 m |V_{min}|} } = 
 \pi \sqrt{2 \Lambda_{Mot}} \; .
\end{equation}
Thus $\Lambda^2_{bec} = 2 \pi^2 \Lambda_{Mot}$ implying that the quantality
inequality in eq.(\ref{e20}) separating solid and mean-field
structures is equivalent to the condition for breakdown of the
classical gas regime in statistical mechanics.

We can evaluate these conditions for the $\alpha-\alpha$ potential $V$
without any bound states parametrized in \cite{ali66} as an attractive
and a repulsive gaussian of different ranges $(r_a,r_r)$ and strengths
$(V_a,V_r)$, i.e.
\begin{eqnarray} \label{e87}
 V(r) = V_r \exp(-r^2/r_r^2) - V_a \exp(-r^2/r_a^2) \; ,
\end{eqnarray}
where the minimum $V_{min}\approx 5-8$~MeV (including the Coulomb
energy of $\approx 2~ $MeV) for $c_{min}\approx 2.5-3.0$~fm and $\Lambda_{Mot}
\approx 0.1-0.2$. Thus $\alpha$-particles would be in the mean-field 
range but with a strong tendency to localize.

\paragraph*{Symmetry requirements.}

Wave functions describing self-bound few-body structures must be
invariant under translations and rotations.  The connection to
conditions for condensate formation is most easily illustrated by use
of an $N$-body wave function $\Psi$ expressed as products of identical
gaussian single-particle wave functions, i.e.
\begin{equation} \label{e33}
 \Psi(\{\bm{r}_i\}) = (b\sqrt{\pi})^{-3N/2}\exp(- \sum_{i=1}^N r_i^2/(2b^2))\;,
\end{equation}
where $\bm{r_{i}}$ is the $i'th$ coordinate.  This as well as all
other mean-field wave functions violate translation invariance, or
equivalently momentum conservation, which is restored by integrating
$\Psi(\{\bm{r}_i-\bm{R}'\}) \exp(i\bm{P}\cdot \bm{R}')$ over all
$\bm{R}'$.  The solution $\Psi_{int}$ of lowest energy has $\bm{P}=0$
which for eq.(\ref{e33}) results in
\begin{eqnarray} \label{e35}
 \Psi_{int}(\{\bm{r}_i\}) = (b\sqrt{\pi})^{-3(N-1)/2} \exp(-\rho^2/(2b^2)) 
  \;,\\  \rho^2 \equiv  \sum_{i=1}^N  q_{i}^2 =
 \frac{1}{N} \sum_ {i<j} r_{ij}^2 = 
 \sum_{i=1}^N r_i^2  - N \bm{R}^2 ,  \label{e36} \\ 
   \bm{q_{i}} \equiv \bm{r_{i}} - \bm{R} \;\;,\;\;
 \bm{r_{ij}} \equiv \bm{r_{i}}-\bm{r_{j}} \;\;,\;\;
 \bm{R} \equiv \frac{1}{N} \sum_{i=1}^N   \bm{r_{i}} \; , \label{e37}
\end{eqnarray}
where the coordinates now are measured from the common center-of-mass
(cm) $\bm{R}$.  This wave function is invariant under rotations around
the cm.

In contrast to the mean-field solution the particles can be correlated
and the wave function $\Psi_{loc}$ in a body-fixed coordinate system
localized in distributions around preferred points $\bm{R}_k$, i.e.
\begin{eqnarray} \label{e43}
\Psi_{loc}(\{\bm{r}_i\})  \propto 
 \sum_{p} \exp(- \sum_{i=1}^N   (\bm{r_{i}} - \bm{R}_{p(i)})^2 /(2B^2)) \;,
\end{eqnarray}
where the normalization is omitted and full symmetry is achieved by
the summation over all permutations $p$ of the set of numbers
$\{1,2,....,N\}$.  The translational invariance is restored precisely
as for eq.(\ref{e33}), i.e. the wave function is obtained from
eq.(\ref{e43}) by the substitution $\bm{r_{i}} \rightarrow \bm{q_{i}}$
in eq.(\ref{e36}) and a corresponding change of normalization
constant.  The rotational invariance is broken for $\Psi_{loc}$ in
eq.(\ref{e43}) but recovered for states of zero angular momentum by
linear combinations of all spatial rotations of $\Psi_{loc}$.

\paragraph*{Condensate assessment.}

To decide if a given wave function describes a condensate we apply
different available definitions. We illustrate again with gaussian
wave functions which overestimates the degree of factorization of the
$N$-body wave function.  One necessary criterion for a condensate is
that the one-body density matrix must have an eigenvalue $\lambda$
comparable in size to $N$ \cite{yang62}.  For a mean-field product
solution the condensate fraction is $c_f=\lambda/N = 1$.  However, the
one-body density matrix is ill-defined for self-bound systems of a
finite number of particles.  This is due to the cm-motion which
decouples completely for correct translationally invariant solutions.
An appropriate cm wave-function could be chosen to allow the usual
definition of the one-body density matrix. The choice could be such
that the condensate fraction $c_f$ is optimized which would be
equivalent to adding an external field as in atomic physics.  For
eq.(\ref{e35}) this recovers the product wave function of all
coordinates in eq.(\ref{e33}) where the cm-motion is completely
ignored.

Instead of using all particle coordinates relative coordinates could
be used and an internal one-body density matrix, $n(\bm{q},\bm{q}')$,
defined \cite{pet00,mat04,suz02,fun03}.  Following \cite{pet00},
i.e. inserting $\bm{q_{N}} = - \sum_{i=1}^{N-1} \bm{q_{i}}$ in
eq.(\ref{e35}), we get
\begin{eqnarray} 
&& n(\bm{q},\bm{q}') \propto \int d^3\bm{q}_2
 d^3\bm{q}_3...d^3\bm{q}_{N-1} |\Psi_{int}|^2 
 \nonumber \\ &\propto&
 \exp\bigg(-\frac{\bm{q}^2 + \bm{q'}^2}{b^2} +
\frac{(N-2)(\bm{q'}+\bm{q})^2}{(N-1) 4 b^2}\bigg) \ \label{e53}\;,
\end{eqnarray}
where $\bm{q}$ and $\bm{q'}$ refer to $\bm{q_{1}}$.  The condensate
fraction obtained through the largest eigenvalue is then \cite{gaj06}
$c_f = 8/(1+\sqrt{2-2/N})^3$ which decreases with $N$ from $1$
for $N=2$ towards about $0.57$ for large $N$.  However, the choice of
internal coordinates is arbitrary \cite{gaj06} and we could as well
choose $\bm{q_{1}}$ supplemented by a set of $N-1$ {\em independent}
Jacobi coordinates.  Then the density matrix corresponding to
eq.(\ref{e35}) would factorize and give $c_f=1$.

For $\alpha$-clusters these options can by appropriate choices lead to
large condensate fractions for rather accurate cluster wave functions.
This is because approximate factorization easily arises at smaller
distances where the potential minimum resembles a harmonic oscillator
and the related $s$-wave solutions resemble gaussians.

Instead of using the eigenvalues of the density matrix a condensate
criterion could be that the one-body (internal) density matrix should
factorize at large distances \cite{yang62}.  This criterion is
extremely difficult to fulfill because the correct nuclear wave
functions never factorize {\em at large distances} as shown in
\cite{mer84}.  Thus, at best only properties at intermediate distances 
could possess condensate properties with this criterion.

Yet another condensate criterion is that all particles occupy the same
quantum state \cite{gaj06}.  This implies that removal of one particle
should leave the single-particle wave functions completely unchanged.
However, for a finite number of particles the remaining interacting
particles would reorganize into a different structure.  This criterion
would be extremely difficult to test directly.

\paragraph*{Localization.}
The $\alpha$-cluster models and the quantality parameter in
eqs.(\ref{e20}) and (\ref{e30}) suggest that localization, or crystal
features of the wave function, may be important.
The resulting condensate fraction depends strongly
on the degree of localization as we can see explicitly by computing
the one-body density matrix for eq.(\ref{e43}).  We assume very narrow
non-overlapping gaussians and an appropriate cm-motion, and obtain
\begin{eqnarray} \label{e45}
 &&n(\bm{r},\bm{r}') = (B\sqrt{\pi})^{-3/2} \\ \nonumber &&\times
\sum_{k=1}^N \exp(- ((\bm{r} - \bm{R}_k)^2 + (\bm{r'} - \bm{R}_k)^2)
  /(2B^2))\;,
\end{eqnarray}
which has $N$ equally large eigenvalues while all others are zero.
This is a condensate fraction of $1/N$ corresponding to one
single-particle state for each of the $N$ particles.  However, after
restoration of rotational symmetry only eigenvalues zero remain.  If
the widths, $B$, of the gaussians increase and they begin to overlap with
each other one eigenvalue separates out and becomes finite.
Increasing the width leads to increasing overlap with a product
wave function like eq.(\ref{e33}). We can quantify by computing the
overlap between the factorized and localized wave functions in
eqs.(\ref{e33}) and (\ref{e43}), i.e.
\begin{eqnarray} \label{e47}
 \langle \Psi | \Psi_{loc} \rangle =  \bigg(\frac{2b B}{b^2+B^2}\bigg)^{3N/2}
 \exp( - \frac{\sum_{k=1}^N {R}_k^2}{2b^2+2B^2})  \;,
\end{eqnarray}
which only is close to unity when $b\sim B$ and either ${R}_i/B \ll 1$
or ${R}_i/b\ll 1$.  Eq.(\ref{e47}) is also obtained by
replacing $\Psi_{loc}$ with the rotationally invariant wave function.
Thus a substantial condensate fraction requires that the overlap with
eq.(\ref{e33}) is large.  However, the spatial extension must be large
to ensure definition (i) of non-overlapping $\alpha$-particles.

\paragraph*{Condensates from cluster models.}

The well-known structure of the Hoyle state in $^{12}$C has about 90\%
overlap with $\alpha$-particles in relative $s$-waves around the center-of-mass
\cite{mat04,suz02,fun03,che07}. This corresponds to an eigenvalue of 
about $0.7$ \cite{mat04,suz02} in agreement with our upper bound of
$c_f=0.80$ derived from eq.(\ref{e35}).  At the same time
$\alpha$-cluster models show $\alpha$-particle density distributions
localized around specific points in space \cite{che07}.  Reconciling
these results, where apparently both localization and large condensate
fraction are present in the same wave function, is only possible with
large widths of the localized wave in eq.(8).  This effectively
recovers the independent particle wave function in eq.(4) where the
$\alpha$-particles are sufficiently separated to remove the need for
nucleon antisymmetrization. These arguments show that the
$\alpha$-condensate states proposed should be regarded as merely an
approximation to existing nuclear $\alpha$-cluster states.

A crucial question is whether a condensate structure can be
experimentally distinguished from other structures. To address this
question computed and measured electron scattering on $^{12}$C was
compared in \cite{che07}.  The conclusions are that $\alpha$-cluster
models of the Brink-type \cite{bri66} and the $\alpha$-condensate
states of \cite{toh01} predict virtually indistinguishable cross
sections and charge distributions.  In addition, these models and
results from more elaborate microscopic calculations \cite{che07} give
precisely the same charge density at large distances.  Thus the
classical cluster parametrization supplemented by nucleon antisymmetry
\cite{toh01} is apparently accurate to about 90\% for the Hoyle state.
However, it is important to realize that fulfilling an ambiguous
definition has very little to do with true condensates which only can
be diagnosed through properties of the wave functions and not by
density distributions. In particular, observable coherence properties
of the many-body states are necessary to separate cluster states from
condensates.  Both this and the localization discussion above strongly
indicate that no new consequences arise from approximating cluster
states with ``condensates''.

\paragraph*{Condensate identity.}

The approximation as a condensate wavefunction of a quantum state
rapidly gets invalid with increasing nuclear mass.  This is seen from
a sequence of four arguments.  First the approximation as a condensate
wavefunction is related to a restricted part of the full Hilbert
space.  Variational computations of condensates assume a class of wave
functions with parameters determined by minimizing the energy.  When
the Hilbert space is extended to include other degrees of freedom the
solution must remain essentially unchanged.  An analogy is found in
the $s$-wave neutron strength function which is broad and distributed
over a large number of many-body states.  This is reflected in the
lack of neutron halo states at excitations around the neutron binding
energy $B_n$ \cite{jen04,jen00}.  To maintain the condensate
character, the residual coupling $V_{c,n}$ of an $\alpha$-condensate
state $|c\rangle$ to the true many-body continuum nuclear states
$|n\rangle$ must all be smaller than their energy difference
\cite{jen00}.

The second argument in the sequence is that this approximation gets
increasingly worse with increasing excitation energy because the
density of states increases.  A "clean" condensate wavefunction must
be more and more "smeared out" over the true many-body states and at
some excitation energy the condensate wavefunction no longer describes
a state of the nucleus.  Third, alpha-condensates are postulated at the
threshold for disintegration into alpha-particles.  Fourth, this is at
an excitation energy $E^*$ of about $7$~MeV for $^{12}$C and
increasing by about $7$~MeV for each additional $\alpha$-particle,
i.e.  $E^* \approx 7 (A/4-2)$~MeV.

We first estimate an average $V_{av}$ of $V_{c,n}$ by using a
nucleon-nucleon potential of range $b$ and strength $V_0$ between the
two states with similar radii $R=bA^{1/3}$.  Both residual kinetic and
potential energy contributions are then proportional to the number of
nucleons $A$, i.e. $|V_{av}| \approx A S_0$.  The radii in $|c\rangle$
and $|n\rangle$ must be comparable if the attraction of short range
has to keep the condensate spatially confined in competition with the
repulsive Coulomb interaction. The condition for maintaining the
condensate character is then $|V_{av}| < D$, where $D$ is the
average level spacing.

We also estimate $V_{av}$ by replacing $|c\rangle$ by the state
$|\alpha,(A-4)\rangle$, consisting of an $\alpha$-particle and the
ground state, $|(A-4)\rangle$, of the $(A-4)$-system. These
wavefunctions are similar, because the condensate consists of
alpha-particles, but they are clearly not identical, since the ground
state of A-4 cannot be accurately described by an alpha-cluster model.
However, they are similar in the approximations of harmonic oscillator
potentials or gaussian wavefunctions.  The differences in the coupling
matrix elements from using $|c\rangle$ and $|\alpha,(A-4)\rangle$ can
then on average be expected to deviate much less than an order of
magnitude.  With many states $|n\rangle$ in the average this leads as
in \cite{sat83,jen00} to the estimate of $V_{av} \approx W_{\alpha}$
where $W_{\alpha}$ is the strength of the imaginary part of the
$\alpha$-nucleus optical potential.  As in \cite{jen00} we conclude
that the condensate resembles one of the many-body states when
$W_{\alpha} < D$.

From the imaginary $\alpha$-nucleus potential we can estimate the
spreading width $\Gamma_{\alpha}$ of an $\alpha$-particle state on the
true many-nucleon states.  For nucleons the spreading width of a
single-particle state of energy $\epsilon$, $\Gamma_{sp}$, is in
Fermi-liquid transport theory \cite{bay76,sie83,hof92} found to be
$\Gamma_{sp} = ((\epsilon-\mu)^2 + \pi^2T^2)/\Gamma_0$ where $\mu$ is
the Fermi level and $T$ the temperature.  The constant $\Gamma_0$ is
related to the imaginary potential and the estimate $\approx 33$~MeV
results in $\Gamma_{sp} \approx 1.5$~MeV for $T=0$ and an energy equal
to the nucleon separation energy \cite{sie83,hof92}.  Analogously we
estimate $\Gamma_{\alpha}$ for $\alpha$-particles moving in the medium
of nucleons.  The $\alpha$-nucleus separation energy is about $7$~MeV,
which with $\Gamma_0 \approx 33$~MeV again results in $\Gamma_{\alpha}
= 1.5 $~MeV for $T=0$.  The finite temperature is obtained from the
average excitation energy, $E^* = a T_c^2$, at the threshold for
fragmentation into free $\alpha$-particles.  With the level density
parameter $a=A/10$MeV we get $T_c \approx 4$~MeV $\sqrt{1-8/A}$ where
$A$ is the nucleon number.  In total we get the estimate
$\Gamma_{\alpha} \approx 6$~MeV almost independent of nucleon
number. This correponds to a strength $W_{\alpha} \approx3$~MeV for
the appropriate increasing excitation energy.

We estimate $D$ in the Fermi gas model adjusted phenomenologically to
excitation energies $E^* \approx B_n$, i.e.  $D
\approx D_0 \exp(-2\sqrt{a(E^*- 2\Delta)})$, where $\Delta \approx
12~$MeV$/\sqrt{A}$ is the pairing gap $\Delta$.  The level spacing,
$D_0 \approx 20~{\rm MeV}/A$, for $E^* = 2\Delta$ is essentially equal
to the single-particle level spacing which is the correct limit.  The
extremely simple expression for $D$ can only be an average over many
nuclei at energies where many excited states are present.

The conditions $|V_{av}| < D$ and $W_{\alpha} < D$ are then
\begin{equation} \label{e80}
 S_{0} A   \,{\rm or}\, W_{\alpha} 
 < D_0 \exp\big(-2\sqrt{a(E^*- 2\Delta)}\big)  \; .
\end{equation}
A very low limit of both $S_0$ and $W_{\alpha}$ is $1$~MeV
\cite{dem02}. With a very small value of $E^* = 2\Delta$ we get the
conservative estimate of preservation of condensate identity $A <
\sqrt{20} < 5$ or $A < 20$, respectively.  The spreading width estimate,
even reduced by a factor of $2$, is also larger than the level
distance for $A < 14$.  These estimates are valid when a sufficient
number of excited states contributes in the average around the
threshold energy. This is fulfilled for all nuclei heavier than
$^{12}$C, including $^{16}$O.  These conditions for survival of the
condensate structures are almost always violated, and the violation
increases exponentially with excitation energy.

\paragraph*{Tonks-Giradeau structures.}

The linear chain structures of $\alpha$-particles at the break-up
threshold, Ikeda diagrams \cite{ike68}, are conceptually similar to
the one-dimensional atomic condensates called Tonks-Giradeau
structures \cite{par04,kin04}. The latter have been realized with
bosonic Rubidium atoms in optical traps that have strong repulsive
interactions in the 1D geometry. This near impenetrability makes the
system behave like a 1D Fermi gas in many aspects. This is analoguous
to impenetrable $\alpha$-particles in 1D cluster structures.  The
corresponding states have been searched for and for many years the
Hoyle state in $^{12}$C was the favorite candidate.  This state is now
claimed as a condensate with a completely different structure.  No
observable has been found to distinguish between these structures
which in any case both are approximated as three-body cluster states.

To assess whether such linear structures could exist in nuclei we turn
to the two conditions in eqs.(\ref{e20}) and (\ref{e30}).  The
$\alpha-\alpha$ Coulomb energy is unimportant compared to $V$ in
eq.(\ref{e87}).  It does not change the condition but it also cannot
provide the confining external field allowing a mean-field
condensate-like solution.  Hence a linear chain structure is only
possible with localized $\alpha$-particles.  The linear chain
structure may also be destroyed by couplings to other degrees of
freedom.  As for three-dimensional condensates we estimate the
survival probability to be very small for any excitation energy above
$2\Delta$.

\paragraph*{Conclusions.}

The existence of Bose-Einstein condensates of $\alpha$-particles
assumes first that nucleons clusterize into $\alpha$-particles, and
second that a condensate is formed.  The quantality condition of
Mottelson indicates that $\alpha$-particles marginally prefer
independent particle motion over correlation. We show that definitions
of condensates of very few particles are ambiguous and lead to
disparate condensate fractions.  The origin is conflicts between
mean-field solutions, correlations and translational or rotational
symmetries, and between definitions related to short and long-range
behavior.  The differences between nuclear and atomic condensates are
few versus macroscopic number of particles, dilute versus high
density, self-bound system versus external confining field, and
ambiguous versus rigorous definitions.

In conclusion, we have found that the concept of a nuclear condensate
is of little use.  The recent theoretical claims of nuclear
$\alpha$-condensates refer to well-known cluster states and can be
regarded as merely an approximation to such states. No observable
differences can be constructed to distinguish these alleged novel
structures from ordinary cluster states.  Occurrence of one and
three-dimensional nuclear condensates of more than 3 particles at
higher excitation energies are very unlikely. 
They would either be completely unstable and vanished into the continuum,
or the $\alpha$-condensate structure would cease to exist due to spreading
over many-nucleon states.  In any case if traces remain they are nothing
else than parts of ordinary $\alpha$-cluster states.

\paragraph*{Acknowledgments.}

Discussions with H.O.U. Fynbo and M. Th\o gersen are highly
appreciated.


\begin{thebibliography}{99}

\bibitem{wig37} E.P. Wigner, Phys. Rev. {\bf 51}, 106 (1937).

\bibitem{whe37} J.A. Wheeler, Phys. Rev. {\bf 52}, 1083 (1937).

\bibitem{wef37} W. Wefelmeier, Naturwiss. {\bf 25}, 525 (1937); 
Z. Phys. {\bf 107}, 332 (1937).

\bibitem{bri66} D. M. Brink, Proc.Int.School Enrico Fermi, course 36,
 Varenna 1965, ed. C. Bloch (Academic Press, 1966).

\bibitem{hoy58} F. Hoyle, Astrophys. J. Suppl. Ser. {\bf 1}, 121 (1954).

\bibitem{cook57} C.W. Cook {\it et al.}, Phys. Rev. {\bf 107}, 508 (1957).

\bibitem{ueg77} E. Uegaki, S. Okabe, Y. Abe, and H. Tanaka,
Prog. Theor. Phys. {\bf 57} (1977) 1262 and {\bf 62}, 1621 (1979).

\bibitem{kan98}  Y. Kanada-En'yo, Phys. Rev. Lett. {\bf 81} (1998) 5291,
and Prog. Theor. Phys. {\bf 117}, 655 (2007).

\bibitem{nef04} T. Neff and H. Feldmeier,
Nucl. Phys. {\bf A 738}, 357 (2004).

\bibitem{des02} P. Descouvemont, Nucl. Phys. {\bf A 709}, 275 (2002).

\bibitem{toh01} Tohsaki {\it et al.},  Phys. Rev. Lett. {\bf 87}, 192501 (2001).

\bibitem{fun02}  Y. Funaki {\it et al.}, Prog. Theor. Phys. {\bf 108}, 297 (2002).

\bibitem{hor70}  H. Horiuchi, Prog. Theor. Phys. {\bf 43}, 375 (1970).

\bibitem{har72} M. Harvey and A.S. Jensen, Nucl. Phys. {\bf A 179} (1972) 33. 

\bibitem{fun03} Y. Funaki {\it et al.}, Phys. Rev. C {\bf 67}, 051306 (2003).

\bibitem{yam04} T. Yamada and P. Schuck, Phys. Rev. C {\bf 69}, 024309 (2004).

\bibitem{ing99} M. Inguscio, S. Stringari, and C.E. Wieman: 
{\em Bose-Einstein condensation in atomic gases}, IOS Press (1999).

\bibitem{dal99} F. Dalfovo {\it et al.}, Rev. Mod. Phys. {\bf 71}, 463 (1999).

\bibitem{mot99} B.R. Mottelson, Nucl. Phys. {\bf A 649}, 45c (1999); and
Proc.Int.School,``Trends in Nuclear Physics, 100 years later'', 
LesHouches 1996.


\bibitem{ali66}  S. Ali and A.R. Bodmer, Nucl. Phys. {\bf 80}, 99 (1966).


\bibitem{mat04} H. Matsumura and Y. Suzuki, Nucl. Phys. {\bf A 739}, 238 (2004).

\bibitem{suz02} Y. Suzuki and M. Takahashi, Phys. Rev. {\bf C 65}, 064318 (2002).

\bibitem{pet00} C.J. Pethick and L.P. Pitaevskii, Phys. 
Rev. {\bf A 62}, 033609 (2000).

\bibitem{gaj06} M. Gajda, Phys. Rev. {\bf A 73}, 023603 (2006).

\bibitem{yang62} C.N. Yang, Rev. Mod. Phys. {\bf 34}, 694 (1962).

\bibitem{mer84} S.P. Merkuriev, S.L. Yakovlev, Teor. Mat. Fiz. {\bf 56}, 60 (1984);
[Theor. Math. Phys. {\bf 56}, 673 (1984)]. 

\bibitem{che07} M. Chernykh {\it et al.}, Phys. Rev. Lett. {\bf 98}, 032501 (2007).

\bibitem{jen04} A.S.~Jensen, K. Riisager, D.V.~Fedorov and E. Garrido,
Rev. Mod. Phys. {\bf 76}, 215 (2004).

\bibitem{jen00} A.S. Jensen and K. Riisager, Phys. Lett. {\bf B 480}, 39 (2000). 

\bibitem{sat83} G.R. Satchler, Direct Nuclear Reactions, Oxford University 
Press, New York 1983, Sections 2.92, p58.

\bibitem{bay76} G.Baym and C.J.Pethick in ``The Physics of Liquid and 
Solid Helium'', K. Bennemann and J. Ketterson (eds.), (Wiley, New York 
1976) 
Vol II. 

\bibitem{sie83} P.J. Siemens,  A.S. Jensen and H. Hofmann, Nucl. Phys.  
{\bf A409}, 135 (1983).

\bibitem{hof92} H. Hofmann, S. Yamaji and A.S. Jensen, Phys. Lett. 
{\bf B286}, 1 (1992).

\bibitem{dem02} P. Demetriou, C. Grama, S. Goriely,  
Nucl. Phys. {\bf A 707}, 253 (2002). 

\bibitem{ike68} K. Ikeda, N. Takigawa and H. Horiuchi
Prog. Theor. Phys. Supplements Extra Number (1968) 464. 

\bibitem{par04} B. Paredes {\it et al.}, Nature {\bf 429}, 277 (2004).

\bibitem{kin04} T. Kinoshita {\it et al.}, Science {\bf 305}, 1125 (2004). 



\end{thebibliography}
\end{document}